# Reconstruction Rating Model of Sovereign Debt by Logical Analysis of Data

Elnaz Gholipour[1*], Béla Vizvári[2], Zoltán Lakner[3]

Declarations

Funding: 'Not applicable'
Conflicts of interest: 'Not applicable'
Availability of data and material: 'Not applicable'
Code availability: 'Not applicable'

[1] Department of Industrial Engineering, Eastern Mediterranean University, Famagusta, North Cyprus Mersin 10 Turkey
* Corresponding Author
E-mail Address; elnaz.gholipour@cc.emu.edu.tr
[2] Department of Industrial Engineering, Eastern Mediterranean University
[3] Department of Food Economics, St. Stephen University

# Abstract

Sovereign debt ratings provided by rating agencies measure the solvency of a country, as gauged by a lender or an investor. It is an indication of the risk involved in investment, and should be determined correctly and in a well-timed manner. The present study reconstructs sovereign debt ratings through logical analysis of data (LAD), which is based on the theory of Boolean functions. It organizes groups of countries according to 20 World Bank-defined variables for the period 2012-2015. The Fitch Rating Agency, one of the three big global rating agencies, is used as a case study. An approximate algorithm was crucial in exploring the rating method, in correcting the agency's errors, and in determining the estimated rating of otherwise non-rated countries. The outcome was a decision tree for each year. Each country was assigned a rating. On average, the algorithm reached almost 98% matched ratings in the training set, and was verified by 84% in the test set. This was a considerable achievement.

# 1. Introduction

## 1.1 Sovereign Credit Ratings (SCRs)

Sovereign debt rating refers to a country 's capability to repay the money that it has borrowed. Therefore, sovereign debt rating can be a metric to help potential investors, financial organizations, banks, and even other governments when making investment or lending decisions with regard to a particular country. Sovereign debt rating reflects the risk involved in doing so. Generally, governments look for a credit ranking to simplify their access to international capital markets, where investors desire to pick the rating securities over unrated ones even with the same credit risk (Cantor and Packer 1996). With an SCR, the government is less reliant on the banks' monetary policy, and it can join international markets. Moreover, SCR can lead to financial improvement by drawing in foreign investors (Luitel and Vanpée 2018).

SCRs play an important role in the credit-rating industry (Hisarciklilar et al. 2011). They can decrease the asymmetric information between investors and borrowers to increase the borrower's willingness to access funds and lessen the credit risk from the lender`s point of view (Canuto, et al. 2012). This process is carried out by credit rating agencies (CRAs) to reduce the information gap between lenders and borrowers. An SCR is supposed to reflect a country's financial, economic, and political position (Kunczik 2002). In particular, since economic and political factors are taken into account by SCRs, it is not easy to measure qualitative variables in the rating procedure in terms of predicting sovereign ratings. Because governments may renege on their obligations, or become less financially solvent due to a political decision, the inclusion of qualitative measures in the rating process is difficult. This is why CRAs give their opinions on the creditworthiness of the country, not investment advice or assessments for obligation (Bozic and Magazzino 2013). This fact was proved by Ferri, Liu, et al. (1999), who noted that CRAs were not able to foresee the East Asia crises of 1999, and this influenced them to become sufficiently conservative to downgrade high-risk countries. Ul Haque et al. (1998) explained the importance of political and economic factors in SCRs. From different point of view, "conflicts of interest" have been used to justify the failures of CRAs (Bernal, Girard, et al. 2016). Arguments abound between experts concerning the effectiveness of these variables on SCRs.

## 1.2 Credit Rating Agencies and Significant Factors

Documented proof is available of how credit rating agencies' shortcomings contributed to the 2008 financial crisis—when they overrated and underrated some countries—and how they can take down governments and blow up capital markets (Mattarocci 2013). At the time in question, they appeared to publish their SCRs in the manner of a black box, while hiding important issues from investors (Gültekin-Karakaş et al. 2011). In the present study, the Fitch agency's rating system has been chosen for analysis. There are a number of financial, political, and social variables that could have been studied when exploring its ratings model. Research on the significant variables of CRAs are used as input, along with the corresponding data from the World Bank, so that the Fitch rating method—otherwise concealed—can be illuminated. Cantor and Packer (1996) were the first researchers in SCR to observe that items such as GDP per capita, GDP growth, inflation, external debt, and default history are important variables. Many other papers have used the same variables while adding new ones. A number of the variables that have been taken into account by different investigators time after time are unemployment, government debt, foreign reserves, fiscal balance, economic development, political stability, mobile phones, real interest rate, total debt, real exchange rate, and unit labor costs (Bissoondoyal-Bheenick et al. 2005), (Philipp et al. 2007), (Bozic and Magazzino 2013). It has been proven that there is a positive impact on government debt and economic growth (Spilioti and Vamvoukas 2015). Alexe et al. (2003) proposed a model for the Standard & Poor agency's rating system by selecting certain economic–financial and political parameters and by using multiple regression. In another study, fiscal uncertainty as a single determinant was used to explain the reason for changes in sovereign ratings during the financial crisis (Hantzsche 2018). The application of the results of previous research requires an expertise in statistics, and is therefore not easy for non-economists.

### 1.3 Overview of the Rating System

Gültekin-Karakaş, Hisarcıklılar et al. (2011) claimed that high-income countries tend to receive higher ratings than low-income ones. They took into consideration some of the same variables, but in their case, GDP per capita was a very important variable for the high-income countries' rating evaluations. When we came to analyse the methodology of previous investigations, we discovered that our technique was both distinctive and untried. Regression analysis using various types of software has been applied in previous investigations (Alexe et al. 2003), (Bozic and Magazzino 2013), (Gültekin-Karakaş et al. 2011), (Philipp, Pedro et al. 2007).

### 1.4 Logical Analysis of Data

Routinely, the regression models that were employed involved different significant indicators that needed to have coefficients as a certain weigh of the related variable to give a country's rating. Therefore, the users of these methods would most probably have had a certain expertise.

Logical analysis of data is a way of studying data sets. It has advantages and disadvantages. The main procedure of LAD is to apply the characteristics of a large and comprehensive set of data from previous monitoring to a much smaller one (Alexe, Alexe et al. 2008). Besides Hammer and Bonates (2006) analysed the application of LAD to medical problems, including the evolution of diagnostic and prognostic systems in cancer investigation. Mirzaei and Vizvári (2011) applied the LAD approach to a one-year series of Moody's ratings. Generally, LAD classifies the outputs as true–false, positive–negative, yes–no or 1–0 to explore the hidden rating processes. While LAD can be applied to science, certain issues have arisen with regard to multiple classes (Avila-Herrera and Subasi 2015).

To sum up, previous studies have revealed drawbacks and advantages in the application of this methodology to CRA policies. The most valuable contribution of the present study is that it offers instructions to all users who wish to rank countries, without them having to possess any special mathematical, financial, or political knowledge. In particular, we have simplified the process by making the cut-points for significant variables of each rank as the output of the logical analysis of data in the form of decision trees.

## 2. Material and Methods

### 2.1 Logical Analysis of Data (LAD)

The present study examined the Fitch agency rating system to reconstruct a rating model of sovereign debts with LAD, a classifying methodology based on optimization and Boolean logic. It uses binary data (i.e., 0 & 1) (Alexe and Hammer 2006). It was initiated by Peter L.

Hammer. He demonstrated that the LAD rating system produces accurate, transparent, and generalizable results (Boros, et al. 1997). There is a set of objects which are somehow similar to each other and are described by the same set of attributes. The objects can be very different according to the area of application; for example, patients in a hospital, customers who obtain a loan from a bank, or drilling locations in the oil industry. However, the nature of the objects is the same in any one application. The objects are divided into two parts, for instance patients who have a particular disease and those who do not. LAD is a method that creates a description for each of the two parts. It is a machine learning method with a supervisor as the classifier of the objects for input into the database. The description thus obtained can be applied to new objects. It is supposed that all attributes of the objects are Boolean variables, that is, the value of each attribute is either true or false. If the values must be expressed numerically, then 1 stands for true and 0 stands for false. It is also assumed that there is no contradiction in the database. There is no pair of objects such that the two objects have the same values in all attributes, but belong to the two aforementioned different subsets of the objects. The number of different objects is at most $2^n$, where $n$ is the number of the binary attributes. However, even in the case when $n$ has a moderate value such as 15 to 20, it is unlikely that all possible observations will have occurred. LAD aims to forecast which new (i.e., until now non-occurring) observation belongs to which subset.

The database can be considered as the description of an incomplete Boolean function of $n$ Boolean variables. The function is incomplete, because its value is not known for all possible values of the attributes (variables), just for the observed values. What LAD must do is to find a complete Boolean function such that its value is exactly the same as the value of the incomplete Boolean function.

The database of the training set must contain a complete classification, that is, each object must be classified as either 1 or 0. LAD can describe any of the two subsets. The synonyms of class 1 and 0 are positive and negative, respectively.

### 2.1.1 Example

The countries are described by economic data that are numeric and not Boolean. These data are transformed in the example given above to Boolean ones. For example, let us consider that GDP per capita is at least $5,436, and this divides countries into two classes. For some, the GDP per capita is $5,436 and above, and it is less in others. A Boolean variable is then introduced. If the condition is satisfied, that is, GDP per capita is at least $5,436, then it is *true* – otherwise a *false* value is obtained It is possible to divide countries into two groups by using the same economic variable in a different way. GDP per capita is used a second time in the example below, where it is at least $14,189. Again, if the condition is satisfied, then it is *true* – otherwise a *false* value is obtained. The transformation of numerical data to Boolean variables is discussed in subsection 2.3.

If a country had a BBBM rating or higher in 2012, then it could be checked in three different ways. If any of the three groups gave a positive result, then the country belonged to that category. If none of the three options were satisfied, then its rating was BBP or worse.

*First group*: The GDP per capita of the country is at least $5,436 AND the export of goods and services is at least 38.185 percentage of GDP AND the PPP conversion factor is at most 6.075%.

*Second group*: The net cash surplus/deficit is at least -5.88 percentage of GDP, that is, the deficit is not too high, AND the total reserves are at least $17,824,012,000 AND the inflation rate of the consumer prices is at most 9.165%

*Third group*: The expenses are not greater than $53.195 AND the male unemployment rate male is at least 8.5% (where the proportion of male labour force is modelled on ILO estimates) AND the GDP per capita is at least $14,189.

To have a rating BBBM or better, a country must satisfy *all* conditions for at least one of the three groups. It is not enough that it satisfies some conditions from every group. Assume for example that the GDP of a country called *Nowhere* is $8,000. The GDP per capita occurs twice, that is, in the 1st and 3rd groups. Nowhere satisfies the GDP constraint of the 1st group, but violates the similar constraints of the 3rd group, that is, overall. Therefore, if its exports are great enough and its PPP conversion factor is low enough, it still can be a BBBM country because it satisfies the criteria of the first group.

### 2.2 The Original Concept of LAD

The original concept of LAD is based on mathematical logic and Boolean variables. One basic theorem of mathematical logic is that every Boolean function can be obtained as a disjunctive normal form (DNF). The three groups of subsection 2.1.1 provide a complete description of the BBBM or better countries. This is an example of a DNF. The three ways consist of conjunctions of Boolean variables. Each of these conjunctions of (perhaps several) Boolean variables are called *patterns* in the context of LAD. The general form of the DNF is that there are several subsets of statements. The Boolean function (the DNF), is true if and only if all statements of at least one subset are true. One statement or its opposite can be a part of several subsets. The subsets of statements may have a different number of elements. It is chance that each group in the example has three statements.

### 2.3 Transformation of Numerical Values to Boolean Attributes

Most real-life problems have numerical attributes, not Boolean ones. LAD is applicable only if the numerical data are "translated" into Boolean attributes. The example of the BBBM or better rating shows how this can be done. Each statement in the example has a numerical value that separates the countries. These numerical values are always between a BBBM or a better country and a country with a lower rating. For instance, of the two countries closest to a GDP of $5,436 per capita, Azerbaijan belonged to the BBBM class in 2012, with a GDP per capita of $7,189. Meanwhile, Guatemala had a lower GDP per capita ($3,166), and was in the BBP class. The statement that the country *Nowhere* has a GDP greater than $5,436 has a Boolean value, as it is either *true* or *false*. The name of any country can be substituted for *Nowhere* in this statement, because every country has a Boolean value in this respect. LAD constructs the DNF from these Boolean attributes. The separating values are called cut-points. Thus, $5,436 is a cut-point in the example above.

Mathematically, binarization can be achieved by introducing cut-points for each of the numerical variables in such a way that the resulting partitioning of space should consist only of "pure intervals," that is intervals that do not contain both positive and negative points (see the red and blue points in Figure 1). Minimizing sets of cut-points with corresponding variables was the optimization element of the present study. We explored and described the rating category of the Fitch rating system by the minimum number of patterns for each year. Figure 1 visualizes the iterative procedure of generating decision trees by logical analysis of data. The countries of the rating category of the iteration and the countries of the better rating classes are assigned to LAD class 1. Any other country is assigned to LAD class 0 like on part" a" of Figure 1. The LAD classes of the countries of the next rating category are changed from 0 to 1 in the next iteration. To go through all pre-defined rating categories, the countries are moved to LAD class 1 gradually with the exception of rating category BM (parts" b" &"c" of Figure 1). At the end, there may be the regions such that countries in the region are not classified by LAD (part" d" of Figure 1). The iterative procedure starts with rating category AAA by Fitch rating agency. These countries are always in LAD class 1. This class is gradually extended by the countries of the other rating categories (AAP, AA, AAM...) in the iterations of the procedure. Moreover, if country not rated by Fitch is in LAD class 1 according to the generated DNF in one of the iterations from AAA to BM, then its suggested rating category is the one when it happens first time. It is an important contribution of this investigation.

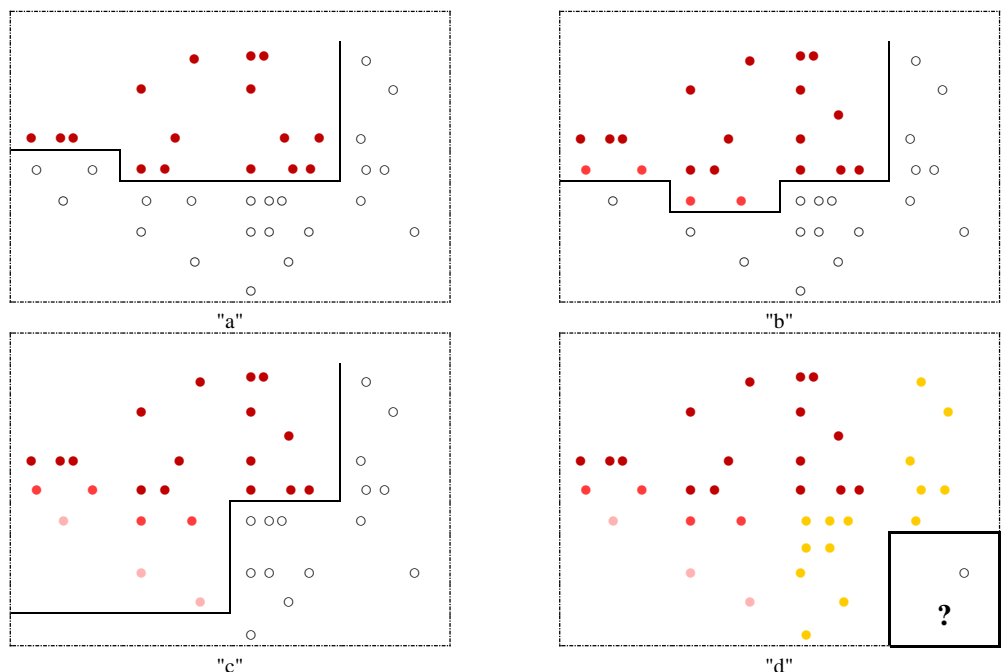

Figure 1: selected cases as an output of LAD for each rating class

## 3. Results

### 3.1 How to apply LAD to sovereign credit rating?

LAD was designed to separate two classes from one another. Sovereign rating has many classes. These classes are ordered according to the risk represented by the countries of the classes. Thus, it is possible to decompose the multi-class rating into a sequence of binary classifications. Countries representing a certain level of risk consist of one class and all other countries with a higher risk are in the other. For example, countries with a rating from AAA to AAM are in the first class and countries having AP or lower are in the second. Every such separation creates a problem for LAD. The DNFs provided by LAD form a decision tree. If a country satisfies the DNF separating AAA countries from other countries, then it is an AAA country. Otherwise, if it satisfies the DNF separating AAA and AAP countries from the others, then it is an AAP country.

### 3.2 Assumptions of LAD Programme and Decision Trees

To apply LAD, we made a number of assumptions. The extent of a pattern is the number of Boolean variables in a conjunction. The prevalence of a positive (negative) pattern is the proportion of positive (negative) observations it covers. The homogeneity of a positive (negative) pattern is the proportion of positive (negative) observations it covers. Every database may require a different value for the LAD parameters. The highest permitted degree for a pattern was 3. The prevalence was at least 70%. The homogeneity was claimed to be 100%. The decision tree for four years from 2012 to 2015 are shown in Tables 3, 4, 5, and 6. The abbreviations of the selected variables which are used in the tables are listed in Table 1.

The data set of 116 countries contains populations of more than half a million in the form of two sets, training, and test: 68 ± 2 and 48 ± 2 countries, respectively. These are gathered from World Bank data covering the period 2012–2015. Countries with very small populations were excluded, as they have a special risk. This fact has been proven by Iceland, who experienced a financial crisis in 2008. The Fitch ratings are listed in Table 2. There are further scales for risky countries. However, the occurrence of these scales is so rare that they are omitted from the calculation.

| Notation | Attribute |
|---|---|
| C | Cash surplus/deficit (%age of GDP) |
| EX | Exports of goods and services (% of GDP) |
| G | GDP per capita (current US$) |
| IM | Imports of goods and services (%age of GDP) |
| RE | Revenue, excluding grants (%age of GDP) |
| SD | Short-term debt (%age of total reserves) |
| TD | Total debt service (%age of exports of goods, services, and primary income) |
| CG | Central government debt, total (%age of GDP) |
| E | Expense (%age of GDP) |
| GG | GDP per capita growth (annual %age) |
| GS | Gross savings (%age of GDP) |
| IV | Industry, value added (%age of GDP) |
| I | Inflation, consumer prices (annual %age) |
| PPP | PPP conversion factor, GDP (LCU per international $) |
| R | Total reserves (includes gold, current US$) |
| U | Urban population (%age of total) |
| PG | Population growth (annual %age) |
| PA | Population ages 0–14 (%age of total) |
| UN | Unemployment, male (%age of male labor force) (modelled ILO estimate) |
| M | Mobile cellular subscriptions (per 100 people) |

Table 1. The applied economic variables

| # | Description | Fitch rating scale |
|---|---|---|
| 1 | (Highest rank): the safest, least risky investments | AAA |
| 2 |  | AAP |
| 3 | (Above Average): safer and less risky | AA |
| 4 |  | AAM |
| 5 |  | AP |
| 6 | (Average): average risk and safety | A |
| 7 |  | AM |
| 8 |  | BBBP |
| 9 | (Slightly below average): slightly risky and less safe | BBB |
| 10 |  | BBBM |
| 11 |  | BBP |
| 12 | (Well below average): riskier and less safe | BB |
| 13 |  | BBM |
| 14 |  | BP |
| 15 | (Lowest): the riskiest and least safe | B |
| 16 |  | BM |

Table 2. Fitch ranking system and descriptions

The following decision trees indicate the prediction of the rating for each country in the set.

### 3.2 Decision Trees

The decision trees obtained from the multiple applications of LAD are as follows:

| Rating Class | Year 2012 |
|---|---|
| | **Decision Tree** |
| AAA | (U>= 73.715, G>=52456.10), OR (U<= 90.02, C>= -9.54, G>=40460.80) |
| AAP | (UN<=3.60, GS>=26.17.5, G>=34424.90), OR(C>=-9.54, U<=90.02, G>=40460.80) |
| AA | (G>=34424.90, CG<=107.355) |
| AAM | (G>=34424.90, CG<=107.355), OR(GS>= 25.435, U>= 73.715, G>=22801.90) |
| AP | (GS>=23.5438, U>=65.98, G>=22293.20), OR(GS>=12.64, U>= 73.03, G>=34424.90), OR(GG>=1.285, PA<=16.985, G>=16463.90), OR (SD<=26.6415, PA>= 15.88, PA<= 18.755) |
| A | (GS>=12.83, U>=84.915, G>=22293.20), OR (CG<=99.96, I>= -0.32, G>=16463.90) , OR(PG>= 0.35, GS>= 22.085, PA<= 19.7.5) |
| AM | (GG >= -1.37, CG<= 109.17, G>=16463.90), OR (CG<= 99.96, E>= 29.4973, U>= 73.03), OR(CG>= 42.545, UN<= 0.91, PA<= 17.785) |
| BBBP | GS>=12.83, GG>= -1.37, G>=15008.10), OR (I<= 5.60, R>= 26351500000, PPP<= 6.075), OR (GS>= 21.18, G>=25449.20, U>= 74.11) |
| BBB | (EX>=39.785, CG<=54.94, G>=8405), OR (GG<=7.92, GS>= 12.83, G>=15008.10), OR(GS>=14.58, E>=16.33, R>=49114000000), OR(M>=91.745, PPP6<= 116.135, R>= 48396000000) |
| BBBM | (G>=5435.98, EX>=38.185, PPP<= 6.075), OR(C>= -5.88, R>= 17824000000, I<= 9.165), OR (E<= 53.195, UN<= 8.5, G>=14189.30) |
| BBP | (R>=47951500000, G>=1357.30), OR(EX>=30.42, GG<=4.36, SD>= 8.625), OR(IV>= 19.845, R>= 3462500000, I<= 6.45 ), OR (PA<=30.175, I<=6.68, GS>=14.04), OR(IV>=21, M>= 106.835, G>=13587.10) |
| BB | C>=-5.75, R>=3462500000, I<=9.165), OR(GS>=14.715, M>= 96, G>=13587.10), OR(GS>=16.51, IV<=31.075, M>= 96), OR (I<= 6.68, IV>= 19.555, PA>= 30.175) |
| BBM | (SD<=23.1219, C>=37.075), OR(GS>=16.925, GG<=4.125, GS<= 30.36), OR(IV>= 21.655, GG<= 5.035, IM>= 26.645) , OR (I<=9.165, IV>=21.095, R>=3462500000), OR(IV>=21.655, PG<= 1.03, PA>= 14.14), OR (PA<= 30.175, I<= 7.905, GS>= 13.505) |
| BP | (EX>=31.19, PG>=0.175, RE>=26.4706), OR (GG<=5.675, GS>=9.27, SD>=10.075), OR(EX>=23.73, GS>=15.78, UN>=0.325) , OR(IV>=21.655, PA<=30.175, M<=121.28)) |
| B | (E<= 53.195, SD<= 23.1219), OR(GS>= 9.27, PPP>= 0.705), OR (GS >= 9.27, UN>= 0.325) |
| BM | (E<= 53.195) |

Table 3 – Decision tree of 2012

| Rating Class | Year 2013 |
|---|---|
| | **Decision Tree** |
| AAA | (G>=47058.30, PA<=19.40), OR(G>=42861.50, EX<= 68.985, GS>= 24.62) |
| AAP | (G>=35760.90, CG<=102.88, GG<=1.515), OR(U>=78.615, M>=122.63, R>=1443740000000) |
| AA | (CG<= 118.525, PPP<=10.545, G>=35760.90), OR(G>=45717.70, U<= 90.575, CG<= 118.52) |
| AAM | (PA<=20.81, CG<= 118.52, G>=23758.60), OR(G>=23029.80, CG<= 63.955, PA >= 14.835) |
| AP | (G>=15203.90, GG>=-0.96, PA<=19.40), OR(IV>= 32.155, I>= 2.805, G>=15172.90), OR (G>=15172.90, PPP>= 0.81, R>= 3240000000), OR(UN>= 3.05, PA<= 21.59, IV>= 35.055) |
| A | (U>=78.615, PPP>=0.505, G>=15203.90), OR (PA<=19.40, GG>=-0.96, G>=15203.90) |
| AM | (U>=72.18, G>=15203.90, PPP>=0/685), OR (PA<=19.40, GG>=0.305, CG>=51.565), OR (CG<= 58.835, IV>= 31.855, EX>= 54.065), OR (UN<= 13.55, I<= 4.02, EX>= 74.515), OR (SD<= 57.89, IV>= 31.85.5, PA >= 19.68) |
| BBBP | (I<=4.24235, R>=43160000000, G>=5808.70), OR (CG<= 50.22, I<= 3.13, G>= 9139.48), OR (I<= 2.685, C>= -3.66, E>=28.825) |
| BBB | (G>=10738.50, I<=5.125, UN<=15.95), OR (IV<= 23.88, TD>= 11.425, CG<= 116.03), OR (IM<= 31.14, GG<= 1.46, EX<= 41.555) |
| BBBM | (PA<= 34.91, CG<= 67.12.71, SD<=68.87), OR (PPP<=6.405, EX>=38.815, G>=5669.61), OR(GS>= 21.57, GS<= 23.945, U<= 94.895) |
| BBP | (TD>=6.66, GS>=18.72, IV<=32.025), OR(G>=5669.61, IV<=39.53, IV>=20.625), OR(GS>=17.485, PA<=29.015, I<=6.735), OR (GG<= 1.46, R>= 4830000000, C>= -3.37) |
| BB | No case |
| BBM | (IV>=22.675, EX>=26.235, GG<=3.91), OR(IV>=25.265, I<=9.835, TD>=10.365), OR(SD>=7.53, I<=5.09, PG>=-0.53), OR (PPP<=266.50, PG<=1.425, IV>=25.265) |
| BP | (UN>=4.35, GS>=13.82, C<=3.65), OR(R>=2505000000, M>=60.91, C>=-8.36) |
| B | (G>=5808.70, GS>=12.985), OR (IV<=39.53, C>=-8.905) |
| BM | No case |

Table 4 – Decision tree of year 2013

| Rating Class | Year 2014 |
|---|---|
| | **Decision Tree** |
| AAA | (CG<= 97.51, G>=44264.40) |
| AAP | (CG<= 97.14, G>=44264.40), OR(IM>= 31.69, R>= 76865000000, G>=37322.70) |
| AA | (GG<= 3.325, CG<= 102.37, G>=37322.70), OR(G>=23895.10, UN <= 7.20, CG<= 76.265) |
| AAM | (PA<=20.59, CG<=117.265, G>=24571) |
| AP | (C<= 0.365, G>=15730.30, E<= 30.185), OR(C>= -4.20, CG<= 102.37, G>=24512.50), OR (E<= 42.62, PG>= 0.465, PA<= 19.10) |
| A | (GG>=-0.725, G>=15730.30, PA<=18.245), OR(GS>=22.90, U>=66.17, GG<=1.985), OR (M<=138.415, I<=2.72, R>=65565000000) |
| AM | (GG>= -0.225, GG<= 1.545, I<= 3.66), OR (M<= 152.09, R>= 6285000000, G>=13680.90), OR(GG>= -0.99, PA<= 19.29, G>=15618.90), OR (GG<= 3.295, GS>= 22.90, U>= 72.715), OR (M<= 138.415, R>= 54060000000, I<= 2.805) |
| BBBP | (I<= 2.91, I>= 0.315, G>=13680.90), OR(R>= 60320000000, UN<= 7.82, PPP<= 14.815), OR(M>= 83.455, U>= 77.645, GG<= 2.285) |
| BBB | R>= 565000000, G>=13680.90, GS>= 17.065), OR(PG>= -0.365, R >= 93800000000, I<= 5.525), OR (UN<= 9.90, U>= 66.505, I<= 2.91) |
| BBBM | (R >= 565000000, G>=13680.90, UN<= 15.10), OR(PG>= 0.15, R>= 32595000000, SD>= 9.635), OR(U>= 54.17, M>= 121.215, PA<= 27.475), OR (UN<= 7.82, PA<= 22.905, M>= 103.16) |
| BBP | (GG<=3.265, TD>=13.52, GS>=20.40), OR(TD>=7.595, GG>=0.195, R>=16430000000), OR (I<=4.345, GG>=-0.03, R>=6935000000), OR(G>=22493.40, I<= 2.76, M>= 83.455) |
| BB | (TD<= 32.585, IM<= 81.065, R>= 12170000000, SD>= 18.68), OR(I>= 0.93, M<= 146.655, M>= 105.905, I5<= 5.66) |
| BBM | (PG <= 1.26, GS>= 13.755, GS<= 3.63), OR(SD>= 32.355, IV>= 23.23, GG<= 4.30), OR (IM<= 53.395, R>= 6285000000, UN<= 7.67), OR (IM<= 48.27, U>= 54.29, EX>= 30.805) |
| BP | (SD>= 32.355, IV>= 16.485, SD<= 59.535), OR(G>= 1873.54, SD>= 32.355, IV>= 21.35), OR(IM>= 31.85, IV>= 28.755, U>= 41.26), OR (GG<= 2.93, I<= 5.455, UN<= 16.45), OR (PG<= 1.485, GS>= 15.50, PG>= -0.325) |
| B | (CG<= 141.95, E>= 31.395), OR(IV>= 11.60, SD>= 0.16, TD<= 24.205) |
| BM | (R<= 455530000000, G<=11380.10), OR(SD>= 0.16, TD<= 35.125) |

Table 5 – Decision tree of year 2014

| Rating Class | Year 2015 |
|---|---|
| | **Decision Tree** |
| AAA | (PA<= 19.675, E<= 42.43, G>=47533.40) |
| AAP | (C<= -1.3747, G>=45329.50, M>= 116.68), OR (UN<= 8.35, PPP<= 10.825, G>=45329.50) |
| AA | (GS>= 24.55, PG>= 0.42, G>=24204.20), OR (GG<= 4.24, PPP<= 10.825, G>=38710.90) |
| AAM | (U<= 87.21, UN<= 6.55, G>=24204.20), OR (GG<= 4.24, PPP<= 10.825, G>=38710.90) |
| AP | (U<= 87.585, G>=17495.60, C>= -1.479), OR(U>= 53.66, GS>= 25.535, R>= 52161000000), OR (PPP<= 10.825, U>= 65.38, G>=38710.90), OR (GG<= 1.655, GG>= 0.74, U>= 79.13) |
| A | (PPP<= 117.435, U>= 65.38, G>=35666.20), OR (C<= 0.305, UN<= 6.20, PA<= 20.835), OR(M>= 117.495, GS>= 23.705, G>=14451.70), OR (M<= 131.47, M>= 114.29, G>=15065) |
| AM | (GS>= 20.755, G>=14179.60, C>= -1.63), OR (UN<= 10.90, R>= 22763500000,G>=14179.60), OR (I<= 2.055, M5>= 111.635, U>= 73.88) |
| BBBP | (EX>= 23.73, C<= -1.245, R>= 82865000000), OR (GS >= 16.94, R>= 8469500000, G>=14179.60), OR (UN <= 5.15, R>= 16708000000, G>=10561.40), OR (M<= 147.075, RE<= 29.38, SD>= 42.33) |
| BBB | (GS>= 19.735, PPP<= 14.99, G>=10327.40), OR(R>= 11365000000, UN<= 5.25, G>=10383.10), OR (I<= 6.34, PG<= 1.345, R>= 44997500000), OR (M<= 134.165, U>= 73.325, M>= 111.15)) |
| BBBM | (PPP<= 1.77, I<= 299, GS>= 14.77), OR(GS>= 23.47, G>=14179.60, PA<= 28.295), OR (IM<= 33.47, M>= 103.755, PA<= 29.275), OR (PA<= 37.02, PPP<= 117.435, R>= 19915000000), OR(U>= 72.05, PG<= 1.06, GS>= 14.77) |
| BBP | (GG>= 0.09, TD>= 10.915, R>= 8469500000), OR(TD>= 5.145, R>= 15619500000, SD>= 11.155), OR(G>= 9845.96, PG<= 1.63, M>= 111.15), OR (PA <= 22.905, GS>= 20.74, M>= 90.31) |
| BB | (GS >= 20.575, M>= 104.745, G>=11039.60), OR(M>= 111.15, PG<= 1.52, G>=10593.80), OR(R>= 66.115, IM<= 47.05, U>= 48.055), OR (PPP<= 5808.54, PG<= 2.295, R>= 15028000000), OR (PA <= 22.905, GS>= 20.575, M>= 104.745) |
| BBM | (I<= 6.80, SD>= 20.235, GG>= 1.335), OR(IV>= 28.14, SD>= 27.36, M>= 105.065), OR (PG<= 1.35, IV>= 28.14, PA>= 16.725), OR (PPP<= 134.13, UN<= 7.61, PPP>= 1.095), OR (PA<= 28.07.5, PPP<= 134.135, GS>= 14.77) |
| BP | (IM<= 47.345, G>= 1889.15, PPP>= 2.585), OR(SD>= 18.905, I<= 7.545, U<= 65.38), OR (PPP<= 45.59, UN<= 19.60, IV<= 27.45), OR(IV>= 17.98, C<= 1.005, PG<= 1.35), OR(TD>= 4.865, I<= 9.515, M>= 111.15) |
| B | (GG>= -4.645, PG>= 1.12), OR (I<= 8.175, R<= 4104500000), OR (C<= 1.005, I<= 38.83, RE<= 44.995) |
| BM | (RE<= 44.995, I<= 38.83, C<= 1.005), OR(PG>= -0.62, IV<= 28.125), OR IV>= 29.46) |

Table 6 – Decision tree of the year 2015

The use of the decision trees is simple. Only the values of the attributes of the country in question must be available. The user must only compare these values with the values of the cut-points given in the table.

### 3.3 Key Variables of the Decision Trees

The four decision trees share certain properties. The most important attributes are listed in table 7.

| Rating Class | Common Variables in all Trees |
|---|---|
| AAA | G |
| AAP – AA – AAM | G |
| AP – A – AM | G – PA – I |
| BBBP – BBB – BBBM | R – GS – G – UN – PPP – I – U |
| BBP – BB – BBM | R – G- GG – SD – IV – I – GS – PG |
| BP – B - BM | UN – IV |

Table 7. The most common properties of the classified ranking classes

The GDP per capita was an extremely important factor in all the rating classes except the lowest. For ratings from AAA to AAM, there was no pattern without it. GDP occurs from AP to AM in more than 50% of the patterns. In addition to GDP, total reserves, gross savings, and inflation are dominant variables in the classification of countries with BBB and BB rating levels. In the lowest rating class, that is B, *male unemployment rates* and *industry value added* are the major variables.

## 4. Discussion

### 4.1 Fitch Rating Predictions

The data set of the study was divided into two subsets, training and test sets. These included 67 and 45 countries respectively. The decision trees are estimated for whole years as well as separately for the two subsets. On average, the decision trees correctly predicted 93%, 89%, 91.5%, 90.5% of all observations (training and test sets) for each of the four years. The high accuracy confirms the value of prognosis and robustness of the LAD rating system. On average, we matched the Fitch rating results in the training set by 98% and in the test set by 84%. The mismatched ratings according to country and the corresponding ratings are shown in Tables 8–11 and the sub-tables.

The year 2012 showed a ratio of matched consequence of 100% for the training set and 85.7% for the test set.

| Country | Our Result | Fitch Rating |
|---|---|---|
| Ukraine | BBBM | B |
| Suriname | BBBM | BBM |
| Rwanda | BP | B |
| Ghana | BB | BP |
| Dominican Republic | BB | B |

Table 8. The mismatched cases of the test set for 2012

The year 2013 showed a ratio of matched consequence of 98.5% for the training set and 78% for the test set. The different output of the two sets are shown in Table 9.

| Country | Our Result | Fitch Rating |
|---|---|---|
| Malta | AP | A |
| Vietnam | BBM | BP |
| Ukraine | BBM | B |
| Kazakhstan | BBB | BBBP |
| Zambia | BP | B |
| Congo, Dem Republic | BBM | BP |
| Paraguay | BBP | BBM |
| Portugal | BBB | BBP |
| Iceland | AAM | BBB |
| Spain | AAM | BBB |
| Cyprus | BBB | B |

Table 9. The mismatched cases of the test set for 2013

| Country | Our Result | Fitch Rating |
|---|---|---|
| Israel | AAM | A |

Table 9-1. The mismatched result of the training set for 2013

For the year 2014, the ratio of matched outcome for training and test sets is 98.5% and 85%. The different output of the two sets is shown below in Tables 10 and 10-1.

| Country | Our Result | Fitch Rating |
|---|---|---|
| South Africa | BBBM | BBB |
| Portugal | A | BBP |
| Panama | B | BBB |
| Iceland | AAM | BBB |
| Cyprus | AP | BM |
| Costa Rica | BBBM | BBP |

Table 10. The mismatched cases of the test set for 2014

| Country | Our Result | Fitch Rating |
|---|---|---|
| Namibia | BBP | BBBM |

Table 10-1. The mismatched result of the training set for 2014

For the year 2015, the ratio of the matched result of training and test sets is 94% and 87%. The unlike rating of two sets are shown in Tables 10 and 11-1.

| Country | Our Result | Fitch Rating |
|---|---|---|
| Congo, Dem Republic | B | BP |
| Rwanda | B | BP |
| Latvia | A | AM |
| Iceland | BBM | BBB |
| Armenia | BBM | BP |

Table 10. The mismatched ratings of the test set for 2015

| Country | Our Result | Fitch Rating |
|---|---|---|
| Ireland | BBB | AM |
| Malaysia | AP | AM |
| Peru | BBB | BBBP |
| Namibia | BBM | BBBM |

Table 11-1. The mismatched ratings of the training set for 2015

First, the high proportions of matched ratings with the Fitch rating system show that our LAD methodology was successful. Second, in the majority of the misclassified cases through the years, Fitch's showed a bias towards downgrading rather than upgrading the countries, as has already been proven by previous studies on CRAs. The incentive in being conservative is that it may deflect criticism, especially in periods of economic and financial crisis. Using our methodology, downgrading outnumbered upgrading considerably – 75.8%, compared with 24.2% in mismatched cases. Fitch had mis rated several countries twice or more than twice during the four years (Table 12).

| Country | No. of misclassification | |
|---|---|---|
| | Twice | More than twice |
| Ukraine | √ | |
| Rwanda | √ | |
| Portugal | √ | |
| Namibia | √ | |
| Iceland | | √ |
| Cyprus | √ | |
| Congo, Dem Republic | √ | |

Table 12. The significant misclassified countries by Fitch between 2012 and 2015

As can be seen, seven countries were mis rated at least twice. The Democratic Republic of Congo, which is rich in minerals, is the main producer of cobalt in the world. There were several military conflicts in the area, however, so the situation was uncertain. Cyprus became a divided country after the events of 1974. No solution has yet been found. The country was also affected by the Greek financial crisis. Iceland was the first victim of the 2008–09 economic crisis. Its population is less than 400,000. Therefore, its currency has also a small total value which was considered by investors to be a source of instability. The Namibian economy is tied to the strong economy of South Africa. Fitch considers Namibia stronger than is merited by its attributes. Portugal is a European Union country in the Mediterranean region. As with other similar countries, it suffered a great deal after the 2008–09 crisis, and accumulated high debts. Rwanda is a neighbor of the Democratic Republic of Congo. Genocide was committed against the country's Tutsi minority in 1994, and the region as a whole remains unstable. Ukraine was another of the countries that were hit by the financial crisis in 2008–09. It is also unstable politically; it has recently gone through a civil war, and it has been in serious conflict with Russia.

## 5. Conclusion

The present study re-constructed the Fitch model of sovereign debt rating using LAD. To do this we examined the behavior of the Fitch rating agency using a World Bank data set containing 20 significant variables for the years 2012–2015. The outcomes of the programme were summarized in the form of the decision trees that can be used by users without the requirement of expertise or without having to have any professional understanding to predict countries' ratings. The decision trees can even be used to estimate the ratings of otherwise unrated countries. This is the main contribution of the present study. The original Fitch ratings and those obtained from the decision trees were in close agreement. Although LAD was designed to separate two classes of objects, it was also helpful in solving the multi-class classification issue.